\documentclass[aps,floatfix,amsmath,nofootinbib,amssymb,superscriptaddress]{revtex4}

\usepackage{overpic}
\usepackage{amssymb}
\usepackage{indentfirst}
\usepackage{feynmf}   
\usepackage{slashed}  
\usepackage{cases}
\usepackage{color}
\usepackage{multirow}
\usepackage{epstopdf}
\usepackage{graphicx,color,bm}
\usepackage{epstopdf}

\usepackage[colorlinks,
            citecolor=green,
            anchorcolor=red,
            menucolor=red,
            linkcolor=red,
            filecolor=red,
            runcolor=red,
            urlcolor=blue,
            frenchlinks=red]{hyperref}

\begin{document}

\title{Gluon generalized parton distributions and angular momentum in a light-cone spectator model}

\author{Chentao Tan}\affiliation{School of Physics, Southeast University, Nanjing
211189, China}

\author{Zhun Lu}
\email{zhunlu@seu.edu.cn}
\affiliation{School of Physics, Southeast University, Nanjing 211189, China}

\begin{abstract}
We study the leading twist gluon generalized parton distributions (GPDs) and the gluon angular momentum inside the proton within a light-cone spectator model. Using the light-cone wave functions derived from the model, we provide the expressions of these distributions at the particular kinematical point $\xi=0$ in the overlap representation. The numerical results of the $H^g$, $E^g$, $\tilde{H}^g$, $H_T^g$ and $E_T^g$ as functions of $x$ at different $\Delta_T$ are presented. Particularly, $H^g$, $\tilde{H}^g$ at non-zero $\Delta_T$ are different from their forward counterparts, the unpolarized distribution $f_1^g$ and the helicity distribution $g_1^g$, respectively. We also obtain the total angular momentum of the gluon contributed to the proton spin $J^g=0.19$, which is consistent with the recent lattice calculation after the uncertainties is considered. The kinetic orbital angular momentum is also calculated and is negative in our model.
\end{abstract}

\maketitle

\section{Introduction}

The study of the hadronic structure in terms of the quark and gluon degrees of freedom is one of the important tasks in hadronic physics.
Although the parton distribution functions(PDFs), which encode the distributions of the longitudinal momentum and polarization carried by quarks and gluons in a fast moving hadron, are most widely used in the investigation of the hadronic structure a more comprehensive picture can be obtained from the general parton distributions(GPDs)~\cite{Muller:1994ses,Diehl:2003ny,Belitsky:2005qn}.
These objects are experimentally accessible through the hard exclusive reactions, such as the deep virtual Compton scattering(DVCS) and deep virtual meson production(DVMP)~\cite{Ji:1996nm,Radyushkin:1997ki,Ji:1998xh,Ji:1998pc,Blumlein:1999sc,Goeke:2001tz}.

One of the necessities of studying GPDs is their relationships with the mass decomposition~\cite{Lorce:2018egm,Hatta:2018sqd} and spin decomposition of hadrons.
In particular, Ji~\cite{Ji:1996ek} derived a gauge-invariant decomposition of the nucleon spin in terms of the quark spin, quark orbital angular momentum (OAM) and the gluon angular momentum.
Furthermore, the angular momentum sum rule can relate the moments of GPDs to the corresponding form factors defined through the expectation value of certain operators, which gives the spin and (orbit) angular momentum of partons evaluated at $t=0$.
The quark OAM and spin-orbit correlations in the nucleon and the pion meson have been calculated by different models~\cite{Lorce:2014mxa,Engelhardt:2021kdo,Lorce:2014kpa,Tan:2021osk}.
In addition, Fourier transforming GPDs with respect to the transverse momentum transfer $\bm{\Delta}_T$ yields the impact-parameter dependent distributions $f(x,\bm{b}_T^2)$~\cite{Burkardt:2000za,Burkardt:2002hr,Bondarenko:2002pp,Riedl:2022pad}, which encode how partons are distributed in the transverse plane along with the longitudinal momentum fraction of partons inside the hadron.

The quark GPDs in the nucleon and the meson have been widely studied from theoretical aspect~\cite{Pasquini:2005dk,Pasquini:2006dv,Meissner:2009ww,Meissner:2008ay,Frederico:2009fk,
Burkardt:2015qoa,Pasquini:2019evu}.
However, the knowledge of the gluon GPDs~\cite{Diehl:2003ny,Polyakov:2002yz} is rather limited.
Nevertheless, the twist-2 gluon GPDs have been calculated in a quark target model~\cite{Meissner:2007rx} which is different from the realistic situation of the target state.
In Ref.~\cite{Meissner:2007rx}, the authors also studied the relations between the gluon GPDs and transverse momentum dependent parton distributions(TMDs).
For $H^g$ and $E^g$, the BFKL resummation technique~\cite{Kuraev:1977fs,Balitsky:1978ic} is established and widely used in phenomenology, and this means that the study on how they evolve mainly focuses on the small-$x$ region~\cite{Gelis:2010nm,Hatta:2022bxn}.
Similarly, the gluon OAM can also be calculated using the sum rule of the corresponding GPDs~\cite{Hatta:2012cs,Kroll:2020jat}, and the methods of extracting it from experimental observables have been proposed~\cite{Goloskokov:2008ib,Ji:2016jgn,Hatta:2016aoc,
Pire:2017yge,Bhattacharya:2017bvs,Bhattacharya:2018lgm,Pire:2021dad,Bhattacharya:2022vvo}.

In this work, we study the leading-twist gluon GPDs and the kinetic gluon OAM from an intuitive model concerning the gluon structure of the proton, which can be considered as a useful complement to the phenomenological analysis on the experimental data and other model calculations.
The approach of the study follows the one applied in Ref.~\cite{Lu:2016vqu}, in which a spectator model was applied to generate the gluon degree of freedom from the proton target to calculate the gluon Sivers function.
In the approach, the proton is regarded as a two-particle system composed by an active gluon and a spectator particle which contains three valence quarks.
Similar model has also been applied to calculate the T-even gluon TMDs~\cite{Bacchetta:2020vty} in which the spectral function and more complicated form factors are considered.
Within the spectator model, we can obtain the expression of the gluon GPDs using the overlap representation~\cite{Diehl:2000xz,Brodsky:2000xy} in terms of the light-cone wave functions of the proton Fock state.
Based on this, we choose the Brodsky-Huang-Lepage prescription~\cite{Brodsky:1982nx} for the coupling of the nucleon-gluon-spectator vertex to calculate the analytical forms of these distributions.
We calculate the five T-even gluon GPDs $H^g$, $E^g$, $\tilde{H}^g$, $H^g_T$ and $E^g_T$ at $\xi=0$.
The GPDs then can be used to study the total angular momentum of the gluon contributed to the proton spin via the Ji's sum rule.
As a byproduct, we also investigate the $x$-dependence of the gluon kinetic OAM.

The rest content of the paper is organized as follows.
In Sec.~\ref{Sec:2}, the definition of the leading-twist gluon GPDs is provided via the light-cone correlation function.
In Sec.~\ref{Sec:3}, we present the analytic expressions of the gluon T-even GPDs in the overlap representation within light-cone spectator model.
In Sec.~\ref{Sec:4}, we provide the results of the GPDs as functions of $x$ at different $\bm{\Delta}_T$. Sec.~\ref{Sec:5}, some conclusions are given.

\section{Definition of the gluon GPDs}\label{Sec:2}

\begin{figure}
  \centering
  \includegraphics[width=0.48\columnwidth]{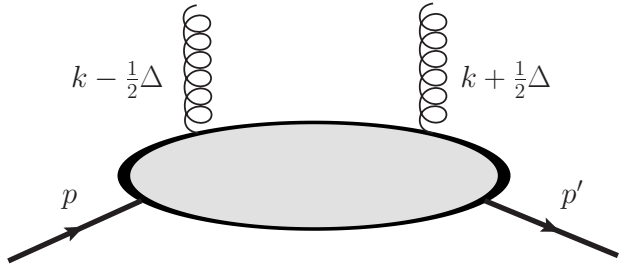}
  \caption{Kinematics for gluon GPDs.}
  \label{fig:gluon}
\end{figure}

In this section, we present the definitions of the gluon GPDs of the proton.
Unless specified otherwise, we will follow the conventions in Ref.~\cite{Diehl:2003ny}. Fig.~\ref{fig:gluon} describes the kinematics for the gluon GPDs. The momenta of the incoming and outgoing proton are given by
\begin{align}
p=P-\frac{1}{2}\Delta, \quad p^\prime=P+\frac{1}{2}\Delta,
\end{align}
where $p^2=p^{\prime 2}=M^2$, with $M$ the proton mass.
In a physical process, $P=(p+p^\prime)/2$ is the average momentum of the initial and final proton, and $\Delta=p^\prime-p$ is the momentum transfer to the proton.
The GPDs depend on three variables
\begin{align}
x=\frac{k^+}{P^+}, \quad \xi=-\frac{\Delta^+}{2P^+}, \quad t=\Delta^2,
\end{align}
where $k$ and $x$ are the average momentum and the average plus-momentum fraction carried by the active gluon, respectively; skewness $\xi$ is the fraction of the transferred momentum; and $t$ is the square of the momentum transfer.
In the light-cone coordinate, a general four vector is defined as
\begin{align}
a^{\pm}=\frac{1}{\sqrt{2}}(a^0 \pm a^3), \quad \bm{a}_T=(a_1,a_2).
\end{align}

Similar to the definitions of PDFs, the GPDs are also defined through the relevant correlation functions. For the leading-twist gluon GPDs, the light-cone correlator reads~\cite{Diehl:2003ny}
\begin{align}
F^{g[ij]}(x,\Delta;\lambda,\lambda^\prime)=\frac{1}{xP^+} \int \frac{dz^-}{2\pi} e^{ik \cdot z} \langle{p^\prime; \lambda^\prime}| F^{+j}_a(-\frac{1}{2}z) \mathcal{W}_{ab}(-\frac{1}{2}z;\frac{1}{2}z) F^{+i}_b(\frac{1}{2}z)|p; \lambda \rangle \big|_{z^+=0^+,\bm{z}_T=\bm{0}_T},
\label{Eq:correlator}
\end{align}
where $\lambda$ and $\lambda^\prime$ denote the helicity of the initial and final proton, respectively. The gluon field strength tensor $F_a^{\mu\nu}(x)$ has the standard form
\begin{align}
F^{\mu\nu}_a(x)=\partial^\mu A^\nu_a(x)-\partial^\nu A_a^\mu(x)+gf_{abc}A_b^\mu(x)A_c^\nu(x),
\end{align}
with $f_{abc}$ being the structure constants of the $SU_c(3)$ group.
To ensure the color gauge invariance of the correlator~(\ref{Eq:correlator}), the Wilson line running along the path
\begin{align}
\mathcal{W}_{ab} \left(-\frac{1}{2}z;\frac{1}{2}z \right) \bigg|_{z^+=0^+,\bm{z}_T=\bm{0}_T}=\left[0^+,-\frac{1}{2}z^-,\bm{0}_T;0^+,\frac{1}{2}z^-,\bm{0}_T\right]_{ab}=\mathcal{P}exp\left[-g\int^{\frac{1}{2}z^-}_{-\frac{1}{2}z^-} dy^- f_{abc} A^+_c(0^+,y^-,\bm{0}_T)\right]
\end{align}
is included, which couples to the gluon field strength tensor through the coupling constant $g$. $\mathcal{P}$ denotes all possible ordered paths followed by the gluon field $A$.
The following two tensors
\begin{align}
\delta^{ij}_T=-g^{ij}, \quad \epsilon^{ij}_T=\epsilon^{-+ij}
\label{eq:tensor}
\end{align}
are applied to define the chiral-even gluon GPDs, while the symmetric operator $\hat{\bm{S}}$
\begin{align}
\hat{\bm{S}}O^{ij}=\frac{1}{2}(O^{ij}+O^{ji}-\delta^{ij}_T O^{mm})
\label{eq:symmetrictensor},
\end{align}
is used to define the chiral-od gluon GPDs, with $O^{ij}$ a general tensor.
Then the twist-2 gluon GPDs can be obtained from the correlator by projecting onto the tensors in Eqs.~\ref{eq:tensor} and (\ref{eq:symmetrictensor}):
\begin{align}
F^g(x,\Delta;\lambda,\lambda^\prime)=&\delta^{ij}_T F^{g[ij]}(x,\Delta;\lambda,\lambda^\prime)= \frac{1}{2P^+}\bar{u}(p^\prime,\lambda^\prime)
\left(\gamma^+H^g(x,\xi,t)+\frac{i\sigma^{+\mu}\Delta_\mu}{2M}E^g(x,\xi,t)\right)u(p,\lambda)
\label{Eq:Fg},\\
\tilde{F}^g(x,\Delta;\lambda,\lambda^\prime)=&i\epsilon^{ij}_T F^{g[ij]}(x,\Delta;\lambda,\lambda^\prime)= \frac{1}{2P^+}\bar{u}(p^\prime,\lambda^\prime)\left(\gamma^+ \gamma_5 \tilde{H}^g(x,\xi,t)+\frac{\Delta^+ \gamma_5}{2M} \tilde{E}^g(x,\xi,t)\right)u(p,\lambda)
\label{Eq:Fgtilde},\\
F^{g,ij}_T(x,\Delta;\lambda,\lambda^\prime)=&-\hat{\bm{S}} F^{g[ij]}(x,\Delta;\lambda,\lambda^\prime)\nonumber \\
=&\frac{\hat{\bm{S}}}{2P^+} \frac{P^+\Delta_T^i-\Delta^+P^i_T}{2MP^+} \bar{u}(p^\prime,\lambda^\prime) \left(i\sigma^{+j}H^g_T(x,\xi,t)+\frac{\gamma^+\Delta_T^j-\Delta^+\gamma^j_T}{2M}E^g_T(x,\xi,t)\nonumber \right.\\
&\left.+\frac{P^+\Delta_T^j-\Delta^+P^j_T}{M^2}\tilde{H}^g_T(x,\xi,t)+\frac{\gamma^+P^j_T-P^+\gamma_T^j}{M}\tilde{E}^g_T(x,\xi,t)\right)u(p,\lambda)
\label{Eq:FgTij}.
\end{align}
Note that Eqs.~(\ref{Eq:Fg},\ref{Eq:Fgtilde}) define the four chiral-even GPDs $H^g$, $E^g$, $\tilde{H}^g$, $\tilde{E}^g$, while Eq.~(\ref{Eq:FgTij}) expresses the  four chiral-odd GPDs $H_T^g$, $E_T^g$, $\tilde{H}_T^g$, $\tilde{E}_T^g$, respectively.
Besides, $E^g$, $H_T^g$, and $\tilde{E}_T^g$ are the helicity-flipped GPDs.

If we choose a particular kinematical point $\xi=0$, which implies the plus-momentum transfer $\Delta^+=0$, the right hand side of Eqs.~(\ref{Eq:Fg}-\ref{Eq:FgTij}) will be simplified considerably.
With the help of the spin vector $S$ of the proton and the transverse component $\bm{\Delta}_T$ of the momentum transfer, we find~\cite{Meissner:2007rx}
\begin{align}
F^g(x,\bm{\Delta}_T;S)=&H^g(x,0,-\bm{\Delta}^2_T)-\frac{i\epsilon^{ij}_T \Delta^i_T S^j_T}{2M}E^g(x,0,-\bm{\Delta}^2_T),\\
\tilde{F}^g(x,\bm{\Delta}_T;S)=&\lambda \tilde{H}^g(x,0,-\bm{\Delta}_T^2),\\
F^{g,ij}_T(x,\bm{\Delta}_T;S)
=&\frac{\hat{\bm{S}}\Delta_T^i\Delta_T^j}{4M^2}\left(E^g_T(x,0,-\bm{\Delta}^2_T)+2\tilde{H}_T^g(x,0,-\bm{\Delta}^2_T)\right) \nonumber \\
&+\frac{i\hat{\bm{S}} \Delta^i_T \epsilon_T^{jk}S_T^k}{2M} \left(H^g_T(x,0,-\bm{\Delta}^2_T)+\frac{\bm{\Delta}^2_T}{4M^2}\tilde{H}_T^g(x,0,-\bm{\Delta}^2_T)\right) \nonumber \\
&-\frac{i\hat{\bm{S}} \Delta^i_T\epsilon_T^{jk}(2\Delta^k_T \bm{\Delta}_T \cdot \bm{S}_T -S^k_T \bm{\Delta}^2_T)}{8M^3}\tilde{H}^g_T(x,0,-\bm{\Delta}^2_T).
\end{align}
Because of the choice $\xi=0$, the GPDs $\tilde{E}^g$ and $\tilde{E}_T^g$ will not show up in the above expressions.

\section{Gluon GPDs in the overlap representation within spectator model}\label{Sec:3}

The gluon GPDs have been first studied by models in Refs.~\cite{Goeke:2006ef,Meissner:2007rx}, in which the authors applied the quark-target model inspired by perturbative QCD, i.e., the gluon is produced from the radiation off the parent quark.
In the case the target is a proton, the minimum Fock state for the proton that containing gluon is $|qqqg\rangle$.
As the four-body system is rather complicated, here we resort to a more phenomenological approach to assume that the three quarks can be grouped into a spectator particle~\cite{Lu:2016vqu,Bacchetta:2020vty}.
Thus, in this model in which the degree of freedom of a gluon is present, the proton can be viewed as a composite system formed by an active gluon and a spectator particle $X$:
\begin{align}
|p;S \rangle \rightarrow |g_{s_g}X_{s_X}(uud)\rangle ,
\end{align}
where $s_g$ and $s_X$ denote the spins of the gluon and the spectator particle, respectively.
In principle the spin quantum number of the spectator can be $s_{\rm{X}}=1/2$ or $3/2$.
Following Ref.~\cite{Lu:2016vqu,Bacchetta:2020vty}, in this work we only consider the spin-1/2 component and neglect the contribution from the spin-$3/2$ component for simplicity.

Then the Fock-state expansion of the proton with $J_z=+1/2$ has the following form:
\begin{align}
|\Psi^\uparrow_{two \ particle}(p^+,\bm{p}_T=\bm{0}_T) \rangle =&\int \frac{d^2\bm{k}_T dx}{16\pi^3\sqrt{x(1-x)}}\nonumber \\
&\times \left[\psi^\uparrow_{+1+\frac{1}{2}}(x,\bm{k}_T)\left|+1,+\frac{1}{2};xp^+,\bm{k}_T \right\rangle+\psi^\uparrow_{+1-\frac{1}{2}}(x,\bm{k}_T)\left|+1,-\frac{1}{2};xp^+,\bm{k}_T \right\rangle \nonumber \right.\\
&\left.+\psi^\uparrow_{-1+\frac{1}{2}}(x,\bm{k}_T)\left|-1,+\frac{1}{2};xp^+,\bm{k}_T \right\rangle+\psi^\uparrow_{-1-\frac{1}{2}}(x,\bm{k}_T)\left|-1,-\frac{1}{2};xp^+,\bm{k}_T \right\rangle \right]
\label{eq:forkstate},
\end{align}
where $\psi^{\uparrow}_{s^z_g s^z_X}(x,\bm{k}_T)$ are the wave functions corresponding to the two-particle states $|s^z_g,s_X^z;xp^+,\bm{k}_T \rangle$, with $s_g^z$ and $s_X^z$ being the $z$ components of the spins of the gluon and spectator, respectively.
Here $\uparrow$($\downarrow$) denotes that the $z$ component $J_z$ of the proton spin $S$ equals $1/2$($-1/2$).
Thus, the light-cone wave functions of the Fock state component of the proton~(\ref{eq:forkstate}) can be expressed as
\begin{align}
\psi^\uparrow_{+1+\frac{1}{2}}(x,\bm{k}_T)&=-\sqrt{2}\frac{-k^1_T+ik^2_T}{x(1-x)}\phi,\nonumber\\
\psi^\uparrow_{+1-\frac{1}{2}}(x,\bm{k}_T)&=-\sqrt{2}\left(M-\frac{M_X}{1-x}\right)\phi,\nonumber\\
\psi^\uparrow_{-1+\frac{1}{2}}(x,\bm{k}_T)&=-\sqrt{2}\frac{+k^1_T+ik^2_T}{x}\phi,\nonumber\\
\psi^\uparrow_{-1-\frac{1}{2}}(x,\bm{k}_T)&=0,
\label{eq:wavefunction+}
\end{align}
which is similar to the light-cone Fock state wave functions of the physical electron given in Ref.~\cite{Brodsky:2000ii}.
Here, $M_X$ is the spectator mass, $\phi$ denotes the wave function in the momentum space
\begin{align}
\phi(x,\bm{k}_T)=\frac{\lambda \sqrt{x} (1-x)}{x(1-x)M^2-(1-x)\bm{k}^2_T-x(\bm{k}_T^2+M_X^2)},
\end{align}
where $M_g$ is the gluon mass for which we fix $M_g=0$, $\lambda$ denotes the coupling of the nucleon-gluon-spectator vertex.
To simulate the nonperturbative physics of the vertex, we choose the Brodsky-Huang-Lepage prescription for the coupling $\lambda$~\cite{Brodsky:1982nx}
\begin{align}
\lambda \rightarrow N_\lambda exp(-\frac{\mathcal{M}^2}{2\beta^2_1}),
\end{align}
where $N_\lambda$ is a strength parameter of the vertex; $\beta_1$ is a cutting-off parameter; and $\mathcal{M}$ is the invariant mass of the two-particle system
\begin{align}
\mathcal{M}^2=\frac{\bm{k}^2_T}{x}+\frac{\bm{k}^2_T+M_X^2}{1-x}.
\end{align}

Similarity, the Fork-state expansion of the proton with $J_z=-1/2$ is given by
\begin{align}
|\Psi^\downarrow_{two \ particle}(p^+,\bm{p}_T=\bm{0}_T) \rangle =&\int \frac{d^2\bm{k}_T dx}{16\pi^3\sqrt{x(1-x)}}\nonumber \\
&\times \left[\psi^\downarrow_{+1+\frac{1}{2}}(x,\bm{k}_T)\left|+1,+\frac{1}{2};xp^+,\bm{k}_T \right\rangle+\psi^\downarrow_{+1-\frac{1}{2}}(x,\bm{k}_T)\left|+1,-\frac{1}{2};xp^+,\bm{k}_T \right\rangle \nonumber \right.\\
&\left.+\psi^\downarrow_{-1+\frac{1}{2}}(x,\bm{k}_T)\left|-1,+\frac{1}{2};xp^+,\bm{k}_T \right\rangle+\psi^\downarrow_{-1-\frac{1}{2}}(x,\bm{k}_T)\left|-1,-\frac{1}{2};xp^+,\bm{k}_T \right\rangle \right],
\end{align}
where
\begin{align}
\psi^\downarrow_{+1+\frac{1}{2}}(x,\bm{k}_T)&=0,\nonumber\\
\psi^\downarrow_{+1-\frac{1}{2}}(x,\bm{k}_T)&=-\sqrt{2}\frac{-k^1_T+ik^2_T}{x}\phi,\nonumber\\
\psi^\downarrow_{-1+\frac{1}{2}}(x,\bm{k}_T)&=-\sqrt{2}\left(M-\frac{M_X}{1-x}\right)\phi,\nonumber\\
\psi^\downarrow_{-1-\frac{1}{2}}(x,\bm{k}_T)&=-\sqrt{2}\frac{+k^1_T+ik^2_T}{x(1-x)}\phi.
\label{eq:wavefunction-}
\end{align}

Similar to the analytical results of the gluon GPDs in the quark target model~\cite{Meissner:2007rx},
we can write the GPDs in the overlap representation using the light-cone wave functions as
\begin{align}
H^g(x,0,-\bm{\Delta}_T^2)=&C_F \underset{s_g^z s_X^z}{\sum}\int \frac{d^2\bm{k}_T}{32\pi^3} \psi^{\uparrow \star}_{s_g^z s_X^z}(x^{out},\bm{k}_T^{out})\psi^\uparrow_{s_g^z s_X^z}(x^{in},\bm{k}_T^{in})+\psi^{\downarrow \star}_{s_g^z s_X^z}(x^{out},\bm{k}_T^{out})\psi^\downarrow_{s_g^z s_X^z}(x^{in},\bm{k}_T^{in})
\label{eq:Hg},\\
\frac{\Delta_-}{2M}E^g(x,0,-\bm{\Delta}_T^2)=&C_F \int \frac{d^2\bm{k}_T}{16\pi^3} \psi^{\uparrow \star}_{+1-\frac{1}{2}}(x^{out},\bm{k}_T^{out})\psi^\downarrow_{+1-\frac{1}{2}}(x^{in},\bm{k}_T^{in})+\psi^{\uparrow \star}_{-1+\frac{1}{2}}(x^{out},\bm{k}_T^{out})\psi^\downarrow_{-1+\frac{1}{2}}(x^{in},\bm{k}_T^{in})
\label{eq:Eg},\\
\tilde{H}^g(x,0,-\bm{\Delta}_T^2)=&C_F \int \frac{d^2\bm{k}_T}{32\pi^3} \psi^{\uparrow \star}_{+1 \pm\frac{1}{2}}(x^{out},\bm{k}_T^{out})\psi^\uparrow_{+1\pm\frac{1}{2}}(x^{in},\bm{k}_T^{in})-\psi^{\uparrow \star}_{-1 +\frac{1}{2}}(x^{out},\bm{k}_T^{out})\psi^\uparrow_{-1+\frac{1}{2}}(x^{in},\bm{k}_T^{in})\nonumber \\
&+\psi^{\downarrow \star}_{-1 \pm\frac{1}{2}}(x^{out},\bm{k}_T^{out})\psi^\downarrow_{-1\pm\frac{1}{2}}(x^{in},\bm{k}_T^{in})-\psi^{\downarrow \star}_{+1 -\frac{1}{2}}(x^{out},\bm{k}_T^{out})\psi^\downarrow_{+1-\frac{1}{2}}(x^{in},\bm{k}_T^{in})
\label{eq:Hgtilde},\\
-\frac{k_-}{M}H^g_T(x,0,-\bm{\Delta}_T^2)=&C_F \int \frac{d^2\bm{k}_T}{16\pi^3} \psi^{\uparrow \star}_{+1-\frac{1}{2}}(x^{out},\bm{k}_T^{out})\psi^\downarrow_{+1-\frac{1}{2}}(x^{in},\bm{k}_T^{in})-\psi^{\uparrow \star}_{-1+\frac{1}{2}}(x^{out},\bm{k}_T^{out})\psi^\downarrow_{-1+\frac{1}{2}}(x^{in},\bm{k}_T^{in})
\label{eq:HgT},\\
-\frac{\Delta_T^1\Delta_T^2}{2M^2}E^g_T(x,0,-\bm{\Delta}_T^2)=&C_F i\int \frac{d^2\bm{k}_T}{16\pi^3} \psi^{\uparrow \star}_{+1+\frac{1}{2}}(x^{out},\bm{k}_T^{out})\psi^\uparrow_{-1+\frac{1}{2}}(x^{in},\bm{k}_T^{in})-\psi^{\uparrow \star}_{-1+\frac{1}{2}}(x^{out},\bm{k}_T^{out})\psi^\uparrow_{+1+\frac{1}{2}}(x^{in},\bm{k}_T^{in})
\label{eq:EgT},\\
\tilde{H}_T^g(x,0,-\bm{\Delta}_T^2)=&0,
\label{eq:HgTtilde}
\end{align}
where $C_F=(N_c^2-1)/(2N_c)$, with $N_c=3$ being the number of the color degree of freedom.
We find that $\tilde{H}_T^g(x,0,-\bm{\Delta}_T^2)$ in Eq.~(\ref{eq:HgTtilde}) vanishes, which is same as the case in the quark target model.
Here the arguments of the initial-state wave functions are given by
\begin{align}
x^{in}&=\frac{x-\xi/2}{1-\xi/2}, \nonumber \\
\bm{k}^{in}_T&=\bm{k}_T- \frac{(1-x)}{1-\xi/2} \frac{\bm{\Delta}_T}{2}
\label{eq:xin},
\end{align}
and those of the final-state wave functions are given by
\begin{align}
x^{out}&=\frac{x+\xi/2}{1+\xi/2}, \nonumber \\
\bm{k}^{out}_T&=\bm{k}_T+ \frac{(1-x)}{1+\xi/2} \frac{\bm{\Delta}_T}{2}
\label{eq:xout}.
\end{align}
Note that Eqs.~(\ref{eq:Hg}-\ref{eq:HgTtilde}) hold at $\xi=0$, so Eqs.~(\ref{eq:xin},\ref{eq:xout}) can be simplified further.

By substituting the light-cone wave functions~(\ref{eq:wavefunction+},\ref{eq:wavefunction-}) into the overlap representation~(\ref{eq:Hg}-\ref{eq:EgT}) of GPDs, we obtain the following analytical results
\begin{align}
H^g(x,0,-\bm{\Delta}_T^2)=&\frac{N_\lambda^2}{24\pi^3x}\int d^2\bm{k}_T exp\left({-\frac{4\bm{k}_T^2+4x M_X^2+(1-x)^2\bm{\Delta}^2_T}{4x(1-x)\beta_1^2}}\right) \nonumber \\
& \frac{4x^2(M_X-M(1-x))^2+(1+(1-x)^2)(4\bm{k}_T^2-(1-x)^2\bm{\Delta}_T^2)}{D^g(x,\bm{\Delta}_T,\bm{k}_T)}
\label{eq:analyticalHg},\\
E^g(x,0,-\bm{\Delta}_T^2)=&-\frac{N_\lambda^2}{3\pi^3} \int d^2\bm{k}_T exp\left({-\frac{4\bm{k}_T^2+4x M_X^2+(1-x)^2\bm{\Delta}^2_T}{4x(1-x)\beta_1^2}}\right)\frac{M(M_X-M(1-x)) (1-x)^2}{D^g(x,\bm{\Delta}_T,\bm{k}_T)}
\label{eq:analyticalEg},\\
\tilde{H}^g(x,0,-\bm{\Delta}_T^2)=&\frac{N_\lambda^2}{24\pi^3}\int d^2\bm{k}_T exp\left({-\frac{4\bm{k}_T^2+4x M_X^2+(1-x)^2\bm{\Delta}^2_T}{4x(1-x)\beta_1^2}}\right) \nonumber \\
&\frac{4x(M_X-M(1-x))^2+(1+(1-x))(4\bm{k}_T^2-(1-x)^2\bm{\Delta}_T^2)}{D^g(x,\bm{\Delta}_T,\bm{k}_T)}
\label{eq:analyticalHgtilde},\\
H^g_T(x,0,-\bm{\Delta}_T^2)=&-\frac{N_\lambda^2}{3\pi^3}\int d^2\bm{k}_T exp\left({-\frac{4\bm{k}_T^2+4x M_X^2+(1-x)^2\bm{\Delta}^2_T}{4x(1-x)\beta_1^2}}\right) \frac{M(M_X-M(1-x))(1-x)}{D^g(x,\bm{\Delta}_T,\bm{k}_T)}
\label{eq:analyticalHgT},\\
E^g_T(x,0,-\bm{\Delta}_T^2)=&-\frac{N_\lambda^2 (1-x)}{3\pi^3x}\int d^2\bm{k}_T exp\left({-\frac{4\bm{k}_T^2+4x M_X^2+(1-x)^2\bm{\Delta}^2_T}{4x(1-x)\beta_1^2}}\right) \frac{\frac{M^2}{\Delta_T^1 \Delta_T^2}(4k_T^1 k_T^2-(1-x)^2\Delta_T^1 \Delta_T^2)}{D^g(x,\bm{\Delta}_T,\bm{k}_T)}
\label{eq:analyticalEgT},
\end{align}
where
\begin{align}
D^g(x,\bm{\Delta}_T,\bm{k}_T)=\left[(\bm{k}_T-\frac{1}{2}(1-x)\bm{\Delta}_T)^2
+xM_X^2-x(1-x)M^2\right]\left[(\bm{k}_T+\frac{1}{2}(1-x)\bm{\Delta}_T)^2+xM_X^2-x(1-x)M^2\right]\, .
\end{align}
We also check that $H^g$ and $\tilde{H}^g$ reduce to the unpolarized PDF $f_1^g(x)$ and helicity PDF $g_1^g(x)$ as $\bm{\Delta}_T \rightarrow 0$.

\section{Numerical results}\label{Sec:4}

In order to present the numerical results of the gluon GPDs, we need to specify the values of parameters $M$, $M_X$, $\beta_1$ and $N_\lambda$ in our model.
We choose~\cite{Lu:2016vqu}
\begin{align}
N_\lambda=5.026, \quad M_X=0.943\ GeV ,\nonumber \\
\beta_1=2.092\ GeV, \quad M=0.938\ GeV
\label{eq:parameter},
\end{align}
which were obtained from a fit of the model result to the leading-order set of the GRV98~\cite{Gluck:1998xa} gluon PDF.
After integrating out $\bm{k}_T$ and substituting the corresponding parameter values in Eqs.~(\ref{eq:analyticalHg}-\ref{eq:analyticalEgT}), we obtain the numerical results of GPDs.

\begin{figure}
  \centering
  \includegraphics[width=0.45\columnwidth]{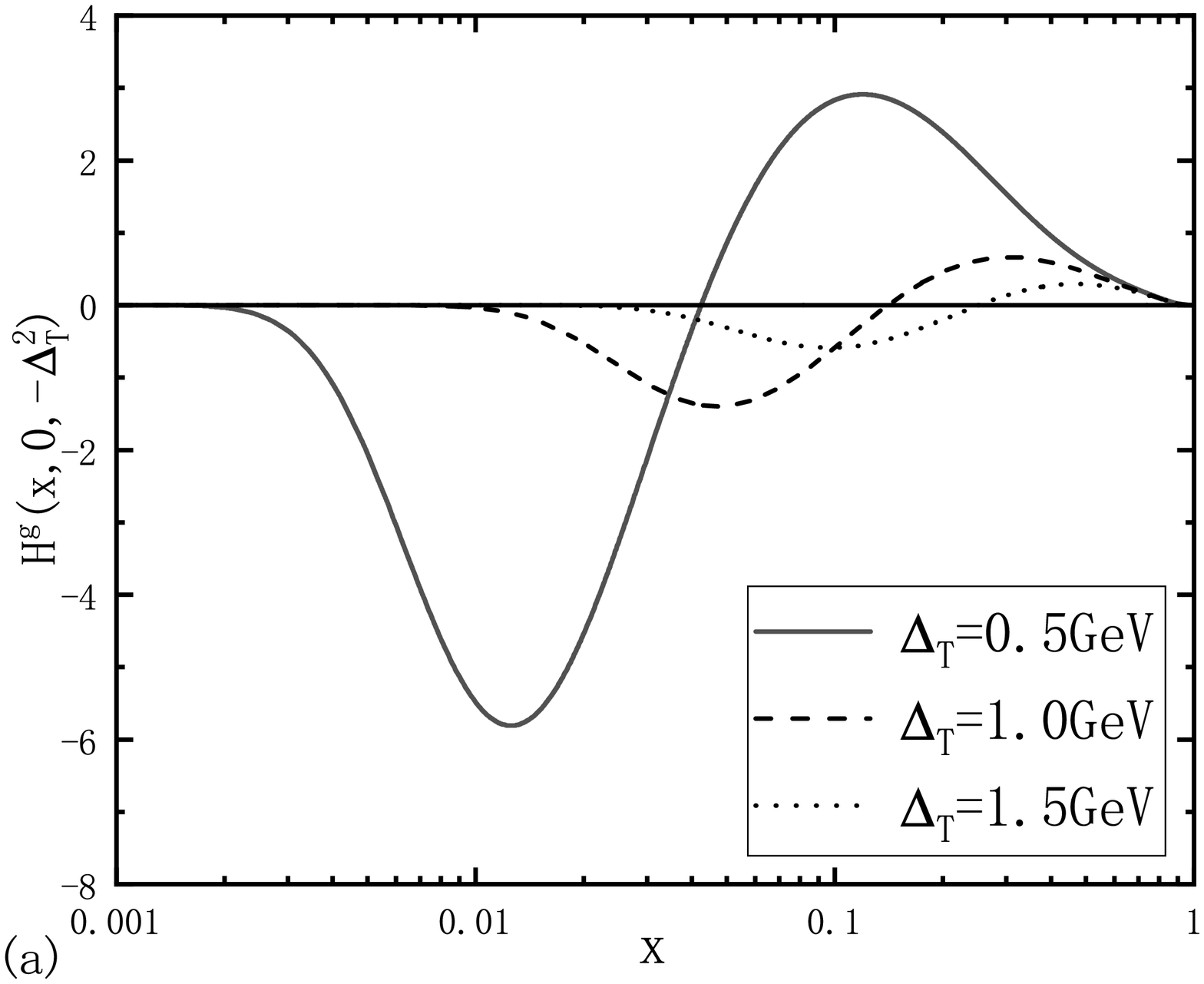}
  \includegraphics[width=0.45\columnwidth]{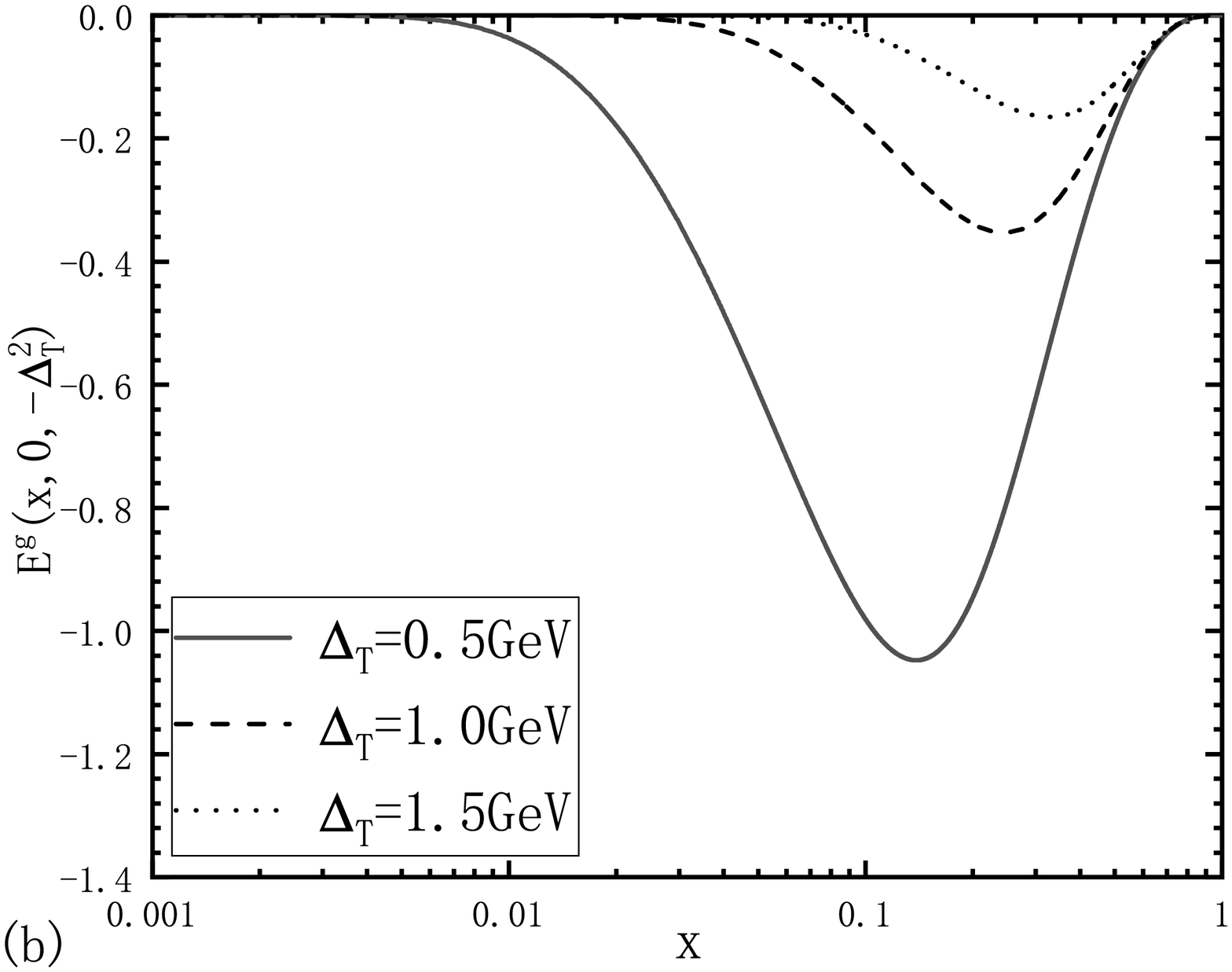}\\
  \includegraphics[width=0.45\columnwidth]{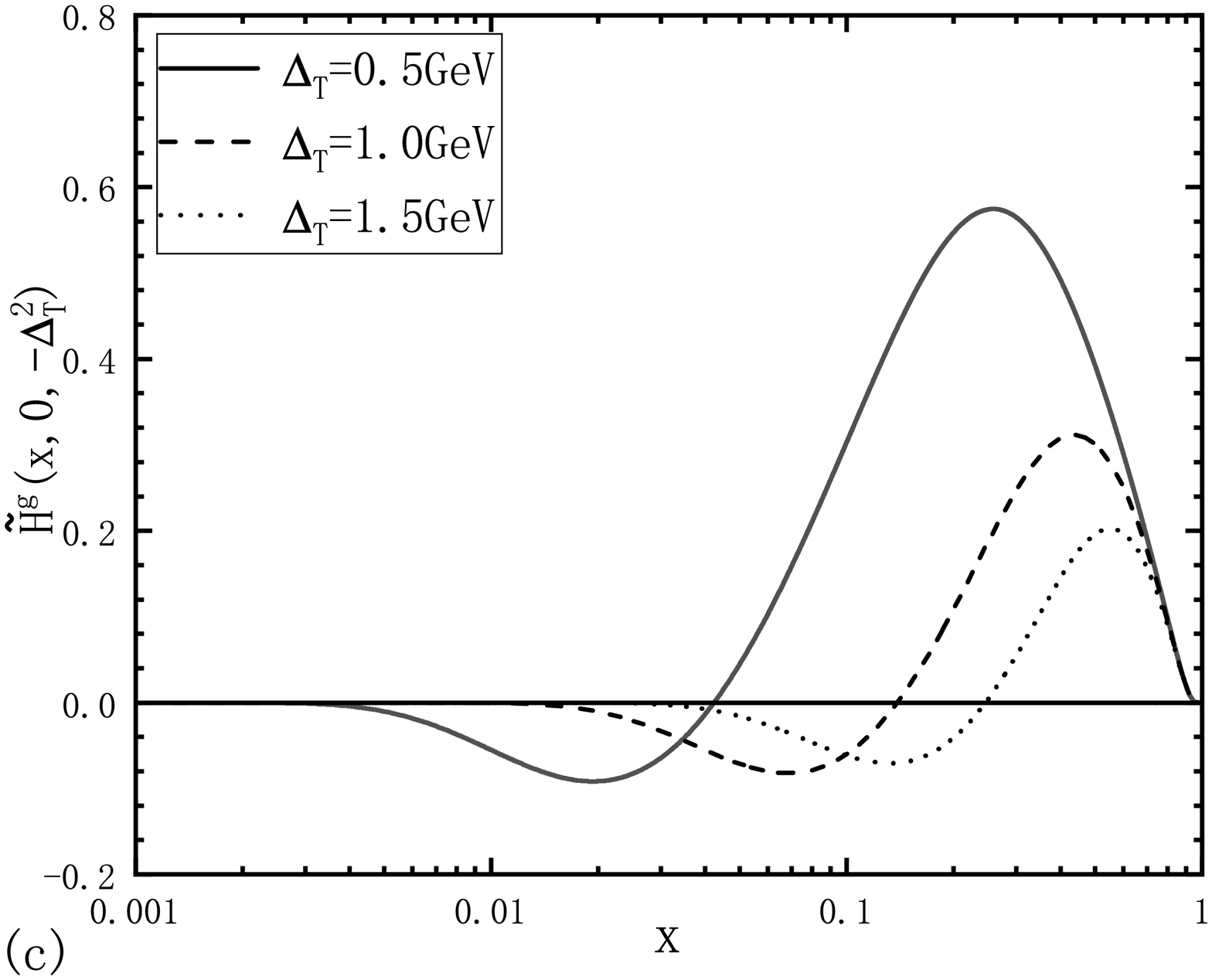}
  \includegraphics[width=0.45\columnwidth]{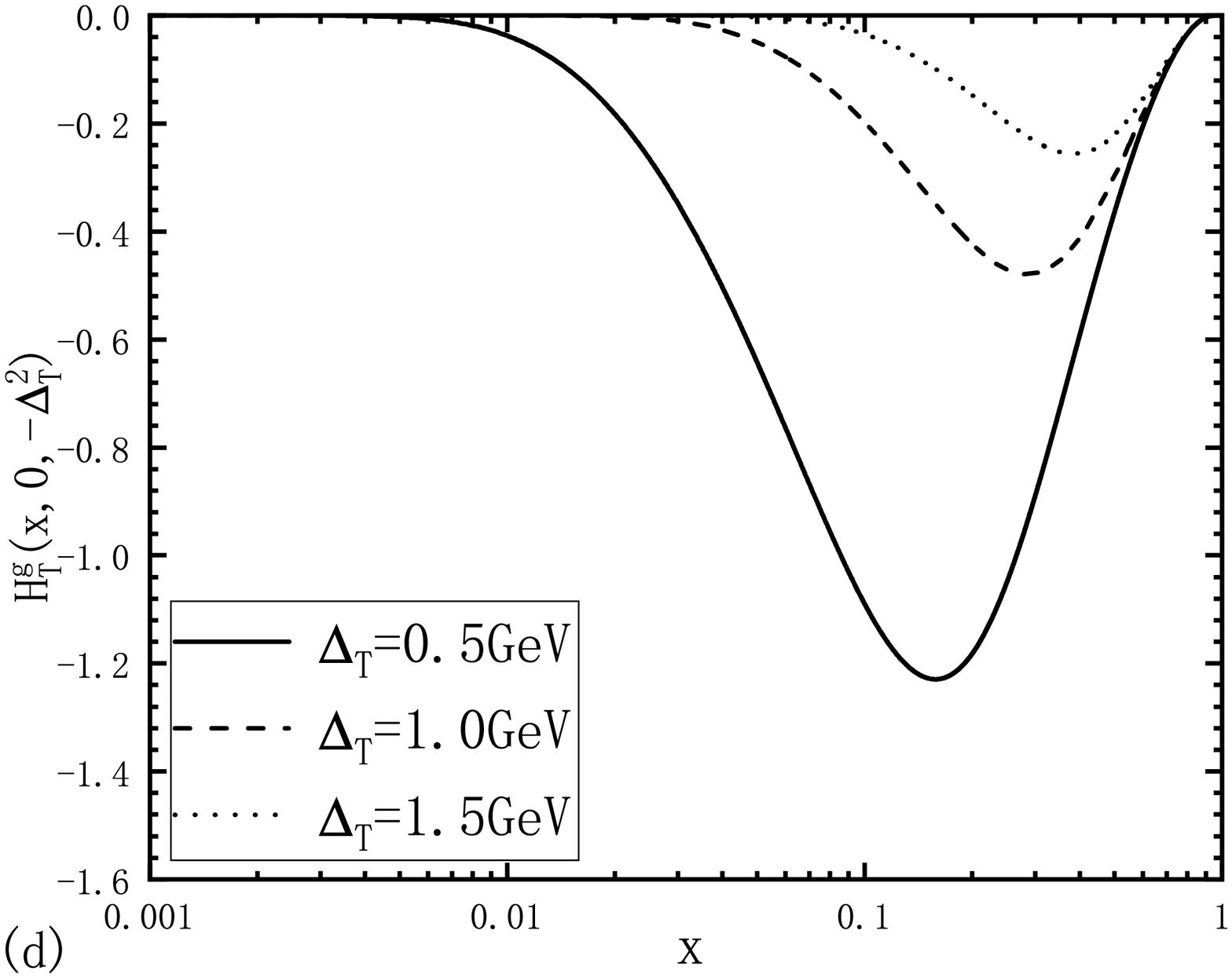}\\
  \includegraphics[width=0.45\columnwidth]{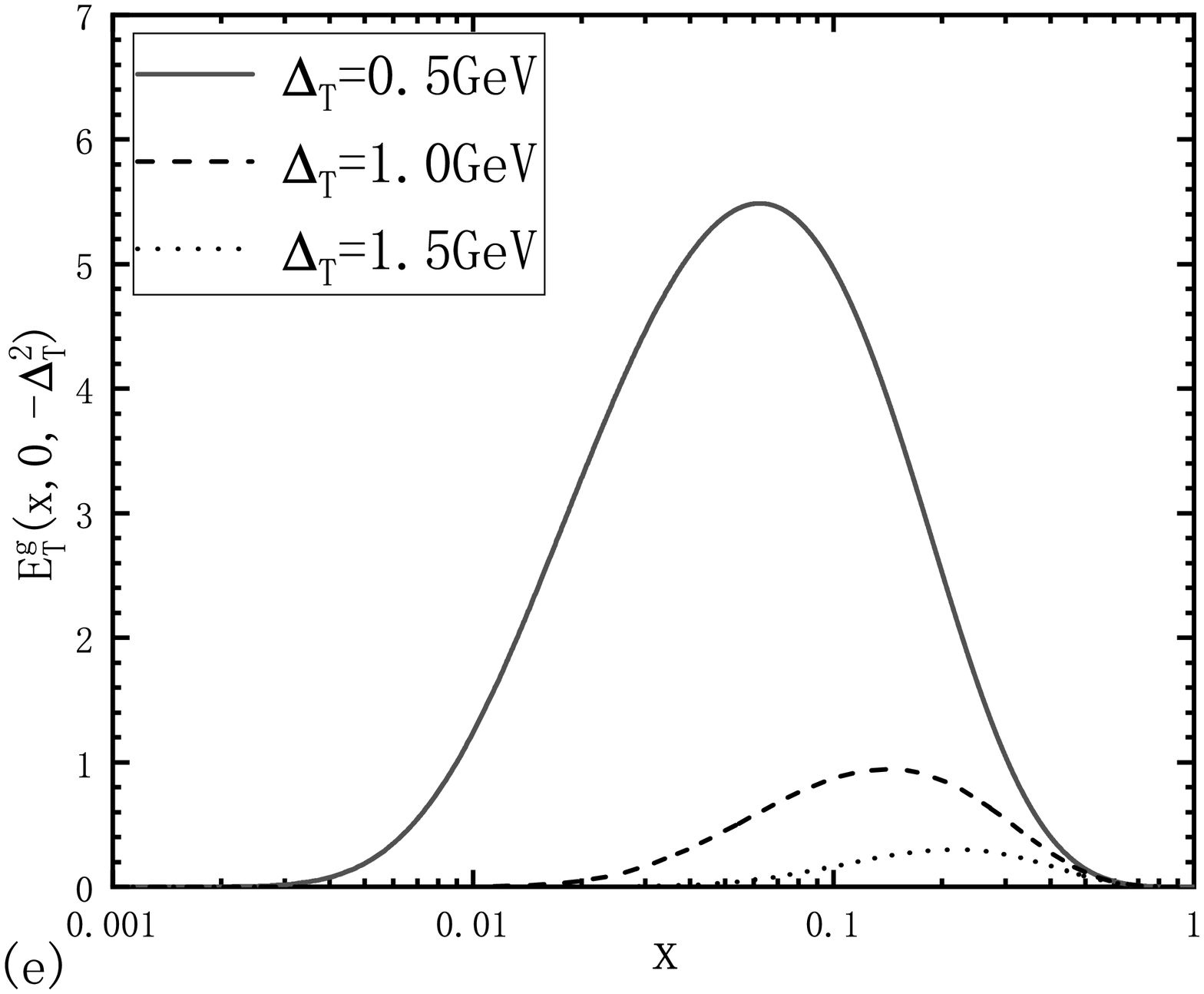}
  \caption{The dependence of the gluon GPDs $H^g(x,0,-\bm{\Delta}_T^2)$, $E^g(x,0,-\bm{\Delta}_T^2)$, $\tilde{H}^g(x,0,-\bm{\Delta}_T^2)$, $H^g_T(x,0,-\bm{\Delta}_T^2)$ and $E^g_T(x,0,-\bm{\Delta}_T^2)$ on $x$ at $\xi=0$ when $\bm{\Delta}_T= 1.0\ GeV,\ 1.5\ GeV,\ 2.0\ GeV$, respectively.}
  \label{fig:GPDs}
\end{figure}

In Fig.~\ref{fig:GPDs}, we plot the GPDs $H^g(x,0,-\bm{\Delta}_T^2)$, $E^g(x,0,-\bm{\Delta}_T^2)$, $\tilde{H}^g(x,0,-\bm{\Delta}_T^2)$, $H^g_T(x,0,-\bm{\Delta}_T^2)$ and $E^g_T(x,0,-\bm{\Delta}_T^2)$ as functions of $x$ at $\bm{\Delta}_T=$0.5 GeV, 1 GeV, 1.5 GeV, respectively.
It is shown that $H^g(x,0,-\bm{\Delta}_T^2)$ has two features which are different from those of the unpolarized gluon PDF $f_1^g(x)$.
First, in the small $x$ region $H^g(x,0,-\bm{\Delta}_T^2)$ is negative, while $f_1^g(X)$ is positive in the whole region $0<x<1$.
That is because there is an additional term $-(1-x)^2\bm{\Delta}_T^2$ in the numerator of the expression (\ref{eq:analyticalHg}).
Second, $H^g(x,0,-\bm{\Delta}_T^2)$ vanishes as $x$ approaches to $0$ when $\Delta_T\neq 0$.
This is different from $f_1^g(x)$ which is nonzero in the small $x$ region.
$H^g(x,0,-\bm{\Delta}_T^2)$ turns to positive in the large $x$ region, i.e., there is a node in the $x$-dependence of $H^g(x,0,-\bm{\Delta}_T^2)$.
Concerning the $\Delta_T$-dependence, the size of $H^g(x,0,-\bm{\Delta}_T^2)$ decreases with increasing $\Delta_T$, and the node position moves toward to higher $x$.

The GPDs $E^g(x,0,-\bm{\Delta}_T^2)$ and $H_T^g(x,0,-\bm{\Delta}_T^2)$ share similar shape since $E^g=x H_T^g$ in our model.
That is, they are both negative in the entire $x$ region, the peak of the curve moves to higher $x$ region with increasing $\Delta_T$.
The GPD $\tilde{H}^g(x,0,-\bm{\Delta}_T^2)$ is negative in the smaller $x$ region and is positive in the larger $x$ region.
Again there is a node in the $x$-dependence of $\tilde{H}^g(x,0,-\bm{\Delta}_T^2)$, which is similar to the case of $H^g(x,0,-\bm{\Delta}_T^2)$.
Finally, the chiral-odd GPD $E_T^g(x,0,-\bm{\Delta}_T^2)$ is positive in the entire $x$ region. It has substantial magnitude in the small x region. while it is largely suppressed in the region $x>0.6$.

The GPDs provide unique opportunity to explore the spin structure of the nucleon.
According to the Ji's sum rule~\cite{Ji:1996ek},
the following moment gives rise to gluon contribution to the nucleon spin:
\begin{align}
J^g=&\int dx \frac{1}{2}\int dx x[H^g(x,0,0)+E^g(x,0,0)]\label{eq:Jg},
\end{align}
We apply our model results for $H^g(x,0,0)$ and $E^g(x,0,0)$ to perform the calculation yielding $$J^g=0.190.$$
This result is consistent with the recent lattice result $J^g=0.186$ calculated by the ETM Collaboration~\cite{Alexandrou:2020sml}.

\begin{figure}
  \centering
  \includegraphics[width=0.5\columnwidth]{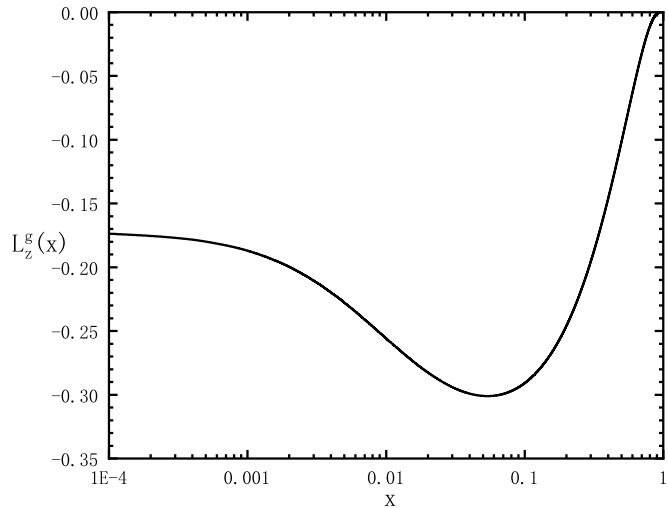}
  \caption{The dependence of the kinetic gluon OAM $L_z^g(x)$ on $x$ in the proton.}
  \label{fig:lzg}
\end{figure}

Using the GPDs $H^g$, $E^g$ and $\tilde{H}^g$, we also calculate the gluon OAM inside the nucleon from the expression~\cite{Ji:1996ek}
\begin{align}
L^g_z=&\frac{1}{2}\int dx \{ x[H^g(x,0,0)+E^g(x,0,0)]-\tilde{H}^g(x,0,0)\}\label{eq:unintegrate}\\
\equiv&\int dx L^g_z(x).\label{eq:lzgx}
\end{align}
This definition corresponds to the kinetic OAM of the gluon~\cite{Chen:2008ag,Wakamatsu:2010qj,Leader:2013jra}, and $L^g_z(x)$ denotes the $x$-dependence of the unintegrated OAM.
Our numerical result shows that $L^g_z=-0.123$,
which means that the gluon kinetic OAM is negative.
In order to present the contribution of the gluon OAM at different $x$, we plot the unintegrated QAM $L_z^g(x)$ as a function of $x$ in Fig.~\ref{fig:lzg}.
We find that $L_z^g(x)$ is negative in the entire $0<x<1$ region.
It is also interesting to point out that the contribution in the very small $x$ region is not zero.
The ``distribution" peaks at around $x=0.05$, and it decreases rapidly when $x$ approaches to 1.
Another observation is that in our model there is substantial cancelation between ${1/2}xH^g(x,0,0)$ and
$\tilde{H}^g(x,0,0)$ because these two GPDs are both positive.
Thus in our result the sign of $L_z^g(x)$ is almost determined by ${1/2}xE^g(x,0,0)$.

\section{Conclusion}\label{Sec:5}

In this work, we studied the leading-twist gluon GPDs of the proton as well as the kinetic gluon OAM by employing a light-cone spectator model.
In the study we took a special kinematic point $\xi=0$.
Thus, among the eight leading-twist GPDs, $H^g$, $E^g$, $\tilde{H}^g$, $H_T^g$, $E_T^g$, and $\tilde{H}_T^g$ survive in this limit.
These GPDs can be expressed as the overlap of the proton wave functions for the in and out states within the light-cone formalism.
In a first approximation we treated the proton as a composite system formed by an active gluon and a spectator particle $X$ to get the proton wave functions.
We chose the Brodsky-Huang-Lepage prescription for the coupling of the proton-gluon-spectator vertex to simulate the nonperturbative physics.

Based on these model assumptions, we obtained the analytic results of the GPDs $H^g$, $E^g$, $\tilde{H}^g$, $H_T^g$ and $E_T^g$.
The GPD $\tilde{H}_T^g$ is found to be zero in this model, similar to the case of the quark GPDs.
We found that in the intermediate $x$ and large $x$ regions $H^g$ and $\tilde{H}^g$ are positive, while in the small $x$ region they are negative and vanishes as $x$ approaches to $0$ and $\Delta_T\neq 0$.
This is different from the PDFs $f_1^g(x)$ and $g_1^g(x)$ which are positive in the whole region $0<x<1$ and is nonzero in the small $x$ region.
Nodes were found in the $x$-dependence of GPDs $H^g$ and $\tilde{H}^g(x,0,-\bm{\Delta}_T^2)$.
It was shown that the GPDs $E^g(x,0,-\bm{\Delta}_T^2)$ and $H_T^g(x,0,-\bm{\Delta}_T^2)$ share similar shape since $E^g=x H_T^g$ in our model, i.e., they are both negative in the entire $x$ region.
The chiral-odd GPD $E_T^g(x,0,-\bm{\Delta}_T^2)$ is positive in the entire $x$ region and has substantial magnitude in the small $x$ region.
Using the Ji's sum rule, we also calculated the total angular momentum of the gluon and obtained $J^g=0.190$, which agrees with recent lattice result within uncertainty.
The kinetic OAM of the gluon in the same model is found to be $L_z^g=-0.123$.
Our study may provide useful theoretical constraints on the gluon GPDs and angular momentum. Further experimental measurements are needed to verify these predictions.

\section*{Acknowledgements}
This work is partially supported by the National Natural Science Foundation of China under grant number 12150013.


\begin{thebibliography}{99}

\bibitem{Muller:1994ses}
D.~M\"uller, D.~Robaschik, B.~Geyer, F.~M.~Dittes and J.~Ho\v{r}ej\v{s}i,
Fortsch. Phys. \textbf{42}, 101-141 (1994).

\bibitem{Diehl:2003ny}
M.~Diehl,
Phys. Rept. \textbf{388}, 41-277 (2003).

\bibitem{Belitsky:2005qn}
A.~V.~Belitsky and A.~V.~Radyushkin,
Phys. Rept. \textbf{418}, 1-387 (2005).

\bibitem{Ji:1996nm}
X.~D.~Ji,
Phys. Rev. D \textbf{55}, 7114-7125 (1997).


\bibitem{Radyushkin:1997ki}
A.~V.~Radyushkin,
Phys. Rev. D \textbf{56}, 5524-5557 (1997).

\bibitem{Ji:1998xh}
X.~D.~Ji and J.~Osborne,
Phys. Rev. D \textbf{58}, 094018 (1998).

\bibitem{Ji:1998pc}
X.~D.~Ji,
J. Phys. G \textbf{24}, 1181-1205 (1998).

\bibitem{Blumlein:1999sc}
J.~Blumlein, B.~Geyer and D.~Robaschik,
Nucl. Phys. B \textbf{560}, 283-344 (1999).

\bibitem{Goeke:2001tz}
K.~Goeke, M.~V.~Polyakov and M.~Vanderhaeghen,
Prog. Part. Nucl. Phys. \textbf{47}, 401-515 (2001).

\bibitem{Lorce:2018egm}
C.~Lorc\'e, H.~Moutarde and A.~P.~Trawi\'nski,
Eur. Phys. J. C \textbf{79}, no.1, 89 (2019).

\bibitem{Hatta:2018sqd}
Y.~Hatta, A.~Rajan and K.~Tanaka,
JHEP \textbf{12}, 008 (2018).

\bibitem{Ji:1996ek}
X.~D.~Ji,
Phys. Rev. Lett. \textbf{78}, 610-613 (1997).

\bibitem{Lorce:2014mxa}
C.~Lorc\'e,
Phys. Lett. B \textbf{735}, 344-348 (2014).

\bibitem{Engelhardt:2021kdo}
M.~Engelhardt, J.~Green, N.~Hasan, T.~Izubuchi, C.~Kallidonis, S.~Krieg, S.~Liuti, S.~Meinel, J.~Negele and A.~Pochinsky, \textit{et al.}
PoS \textbf{LATTICE2021}, 413 (2022).

\bibitem{Lorce:2014kpa}
C.~Lorc\'e,
Int. J. Mod. Phys. Conf. Ser. \textbf{37}, 1560036 (2015).

\bibitem{Tan:2021osk}
C.~Tan and Z.~Lu,
Phys. Rev. D \textbf{105}, no.3, 034004 (2022).

\bibitem{Burkardt:2000za}
M.~Burkardt,
Phys. Rev. D \textbf{62}, 071503 (2000).

\bibitem{Burkardt:2002hr}
M.~Burkardt,
Int. J. Mod. Phys. A \textbf{18}, 173-208 (2003).

\bibitem{Bondarenko:2002pp}
S.~Bondarenko, E.~Levin and J.~Nyiri,
Eur. Phys. J. C \textbf{25}, 277-286 (2002).



\bibitem{Riedl:2022pad}
C.~Riedl,
doi:10.5506/APhysPolB.53.5-A2
[arXiv:2204.03684 [hep-ex]].

\bibitem{Pasquini:2005dk}
B.~Pasquini, M.~Pincetti and S.~Boffi,
Phys. Rev. D \textbf{72}, 094029 (2005)
[arXiv:hep-ph/0510376 [hep-ph]].

\bibitem{Pasquini:2006dv}
B.~Pasquini and S.~Boffi,
Phys. Rev. D \textbf{73}, 094001 (2006)
[arXiv:hep-ph/0601177 [hep-ph]].

\bibitem{Meissner:2009ww}
S.~Meissner, A.~Metz and M.~Schlegel,
JHEP \textbf{08}, 056 (2009).

\bibitem{Meissner:2008ay}
S.~Meissner, A.~Metz, M.~Schlegel and K.~Goeke,
JHEP \textbf{08}, 038 (2008).

\bibitem{Frederico:2009fk}
T.~Frederico, E.~Pace, B.~Pasquini and G.~Salme,
Phys. Rev. D \textbf{80}, 054021 (2009)
[arXiv:0907.5566 [hep-ph]].

\bibitem{Burkardt:2015qoa}
M.~Burkardt and B.~Pasquini,
Eur. Phys. J. A \textbf{52}, no.6, 161 (2016)
[arXiv:1510.02567 [hep-ph]].

\bibitem{Pasquini:2019evu}
B.~Pasquini, S.~Rodini and A.~Bacchetta,
Phys. Rev. D \textbf{100}, no.5, 054039 (2019)
[arXiv:1907.06960 [hep-ph]].

\bibitem{Polyakov:2002yz}
M.~V.~Polyakov,
Phys. Lett. B \textbf{555}, 57-62 (2003).

\bibitem{Meissner:2007rx}
S.~Meissner, A.~Metz and K.~Goeke,
Phys. Rev. D \textbf{76}, 034002 (2007).

\bibitem{Kuraev:1977fs}
E.~A.~Kuraev, L.~N.~Lipatov and V.~S.~Fadin,
Sov. Phys. JETP \textbf{45}, 199-204 (1977).

\bibitem{Balitsky:1978ic}
I.~I.~Balitsky and L.~N.~Lipatov,
Sov. J. Nucl. Phys. \textbf{28}, 822-829 (1978).

\bibitem{Gelis:2010nm}
F.~Gelis, E.~Iancu, J.~Jalilian-Marian and R.~Venugopalan,
Ann. Rev. Nucl. Part. Sci. \textbf{60}, 463-489 (2010).

\bibitem{Hatta:2022bxn}
Y.~Hatta and J.~Zhou,
[arXiv:2207.03378 [hep-ph]].

\bibitem{Kroll:2020jat}
P.~Kroll,
Mod. Phys. Lett. A \textbf{35}, no.12, 2050093 (2020)
[arXiv:2001.01919 [hep-ph]].

\bibitem{Hatta:2012cs}
Y.~Hatta and S.~Yoshida,
JHEP \textbf{10}, 080 (2012).

\bibitem{Goloskokov:2008ib}
S.~V.~Goloskokov and P.~Kroll,
Eur. Phys. J. C \textbf{59}, 809-819 (2009)
[arXiv:0809.4126 [hep-ph]].

\bibitem{Ji:2016jgn}
X.~Ji, F.~Yuan and Y.~Zhao,
Phys. Rev. Lett. \textbf{118}, no.19, 192004 (2017).

\bibitem{Hatta:2016aoc}
Y.~Hatta, Y.~Nakagawa, F.~Yuan, Y.~Zhao and B.~Xiao,
Phys. Rev. D \textbf{95}, no.11, 114032 (2017).

\bibitem{Pire:2017yge}
B.~Pire and L.~Szymanowski,
Phys. Rev. D \textbf{96}, no.11, 114008 (2017)
[arXiv:1711.04608 [hep-ph]].




\bibitem{Bhattacharya:2017bvs}
S.~Bhattacharya, A.~Metz and J.~Zhou,
Phys. Lett. B \textbf{771}, 396-400 (2017)
[erratum: Phys. Lett. B \textbf{810}, 135866 (2020)].

\bibitem{Bhattacharya:2018lgm}
S.~Bhattacharya, A.~Metz, V.~K.~Ojha, J.~Y.~Tsai and J.~Zhou,
[arXiv:1802.10550 [hep-ph]].

\bibitem{Pire:2021dad}
B.~Pire, L.~Szymanowski and J.~Wagner,
Phys. Rev. D \textbf{104}, no.9, 094002 (2021)
[arXiv:2104.04944 [hep-ph]].

\bibitem{Bhattacharya:2022vvo}
S.~Bhattacharya, R.~Boussarie and Y.~Hatta,
Phys. Rev. Lett. \textbf{128}, no.18, 182002 (2022).

\bibitem{Lu:2016vqu}
Z.~Lu and B.~Q.~Ma,
Phys. Rev. D \textbf{94}, no.9, 094022 (2016).

\bibitem{Bacchetta:2020vty}
A.~Bacchetta, F.~G.~Celiberto, M.~Radici and P.~Taels,
Eur. Phys. J. C \textbf{80}, no.8, 733 (2020).

\bibitem{Diehl:2000xz}
M.~Diehl, T.~Feldmann, R.~Jakob and P.~Kroll,
Nucl. Phys. B \textbf{596}, 33-65 (2001)
[erratum: Nucl. Phys. B \textbf{605}, 647-647 (2001)]
[arXiv:hep-ph/0009255 [hep-ph]].

\bibitem{Brodsky:2000xy}
S.~J.~Brodsky, M.~Diehl and D.~S.~Hwang,
Nucl. Phys. B \textbf{596}, 99-124 (2001)
[arXiv:hep-ph/0009254 [hep-ph]].


\bibitem{Brodsky:1982nx}
S.~J.~Brodsky, T.~Huang and G.~P.~Lepage,
Springer Tracts Mod. Phys. \textbf{100}, 81-144 (1982)
SLAC-PUB-2868.

\bibitem{Goeke:2006ef}
K.~Goeke, S.~Meissner, A.~Metz and M.~Schlegel,
Phys. Lett. B \textbf{637}, 241-244 (2006).

\bibitem{Brodsky:2000ii}
S.~J.~Brodsky, D.~S.~Hwang, B.~Q.~Ma and I.~Schmidt,
Nucl. Phys. B \textbf{593}, 311-335 (2001).

\bibitem{Gluck:1998xa}
M.~Gl\"uck, E.~Reya and A.~Vogt,
Eur. Phys. J. C \textbf{5}, 461-470 (1998).

\bibitem{Alexandrou:2020sml}
C.~Alexandrou, S.~Bacchio, M.~Constantinou, J.~Finkenrath, K.~Hadjiyiannakou, K.~Jansen, G.~Koutsou, H.~Panagopoulos and G.~Spanoudes,
Phys. Rev. D \textbf{101}, no.9, 094513 (2020).

\bibitem{Chen:2008ag}
X.~S.~Chen, X.~F.~Lu, W.~M.~Sun, F.~Wang and T.~Goldman,
Phys. Rev. Lett. \textbf{100}, 232002 (2008)
[arXiv:0806.3166 [hep-ph]].

\bibitem{Wakamatsu:2010qj}
M.~Wakamatsu,
Phys. Rev. D \textbf{81}, 114010 (2010)
doi:10.1103/PhysRevD.81.114010
[arXiv:1004.0268 [hep-ph]].

\bibitem{Leader:2013jra}
E.~Leader and C.~Lorc\'e,
Phys. Rept. \textbf{541}, no.3, 163-248 (2014)
[arXiv:1309.4235 [hep-ph]].

\end{thebibliography}
\end{document}